\documentclass[a4paper, iop, twocolumn, twocolappendix, numberedappendix, appendixfloats]{emulateapj}

\pdfoutput=1
\usepackage[breaklinks,colorlinks=true,linkcolor=blue,anchorcolor=blue,citecolor=blue,urlcolor=blue,draft=false]{hyperref}
\usepackage{color, graphicx, amssymb, mathrsfs, placeins, times, xspace}

\usepackage{natbib}
\usepackage[utf8]{inputenc}

\newcommand{\percent}{\ensuremath{\%}}
\newcommand{\Msun}{M_\odot}

\newcommand{\Mbar}{M_\mathrm{bar}}
\newcommand{\gbar}{\mathrm{g}_\mathrm{bar}}
\newcommand{\gobs}{\mathrm{g}_\mathrm{obs}}
\newcommand{\gdag}{\mathrm{g}_\mathrm{\dagger}}
\newcommand{\gddag}{\mathrm{g}_\mathrm{\ddagger}}
\newcommand{\spaxel}{\texttt{Spaxel}}
\newcommand{\marvin}{\texttt{Marvin}}

\shorttitle{MVDR in BCGs}
\shortauthors{Tian et al.}

\begin{document}
\title{Mass-Velocity Dispersion Relation in MaNGA Brightest Cluster Galaxies}
\author{Yong Tian\altaffilmark{1}}
\author{Han Cheng\altaffilmark{1}}
\author{Stacy S. McGaugh\altaffilmark{2}}
\author{Chung-Ming Ko\altaffilmark{1,3,$\dagger$}}
\author{Yun-Hsin Hsu\altaffilmark{4,5}}

\email{cmko@astro.ncu.edu.tw}
\email{stacy.mcgaugh@case.edu}
\altaffiltext{1}{Institute of Astronomy, National Central University, Taoyuan 32001, Taiwan}
\altaffiltext{2}{Department of Astronomy, Case Western Reserve University, 10900 Euclid Avenue, Cleveland, Ohio 44106, USA}
\altaffiltext{3}{Department of Physics and Center for Complex Systems, National Central University, Taoyuan 32001, Taiwan}
\altaffiltext{4}{Academia Sinica Institute of Astronomy and Astrophysics (ASIAA), No. 1, Section 4, Roosevelt Road, Taipei 10617, Taiwan}
\altaffiltext{5}{Institute of Astronomy, National Tsing Hua University, No. 101, Section 2, Kuang-Fu Road, Hsinchu 30013, Taiwan}
\altaffiltext{$\dagger$}{Corresponding author}

%
%
\begin{abstract}
We investigate a kinematic scaling relation between the baryonic mass and
 the flat velocity dispersion, i.e. mass-velocity dispersion relation (MVDR),
 from the brightest cluster galaxies (BCGs) to the galaxy clusters.
In our studies, the baryonic mass of BCGs is mainly estimated by photometry.
The velocity dispersion profiles are explored with the integrated field unit (IFU)
 by Mapping Nearby Galaxies at Apache Point Observatory (MaNGA).
For the first time, we reveal two significant results with 54 MaNGA BCGs:
(1) the flat velocity dispersion profiles;
(2) a tight empirical relation on the BCG-cluster scale together with cluster samples, i.e., MVDR,
 $\log(M_\mathrm{bar}/M_\odot)=4.1^{+0.1}_{-0.1}\log(\sigma_{\mathrm{los}}/\mathrm{km}\,\mathrm{s}^{-1})
 +1.6^{+0.3}_{-0.3}$, with a tiny lognormal intrinsic scatter of $10^{+2}_{-1}\%$.
This slope is identical to the acceleration relation in galaxy clusters,
 which is reminiscent of the spiral galaxies,
 albeit at a larger characteristic acceleration scale.
The residuals of the MVDR represent a Gaussian distribution,
 displaying no correlations with four properties:
 baryonic mass, scale length, surface density, and redshift.
Notably, the MVDR on the BCG-cluster scale provides a strict test,
 which disfavors the general prediction of the slope of three in the dark matter model.
\end{abstract}

\keywords{Galaxy kinematics (602); Brightest cluster galaxies (181); Giant elliptical galaxies (651);  Galaxy clusters (584); Dark matter (353)}

%
%

\section{Introduction}\label{sec:intro}
A kinematic scaling relation is the counterpart of a dynamical scaling relation,
 which plays a major role in understanding fundamental physics.
For examples, Kepler's law leads to Newtonian dynamics in the Solar system.
However, Kepler's law is no longer applicable in spiral galaxies,
  well-known as the dark matter (DM) problem.
Instead, a different kinematic relation was discovered between the baryonic mass and
 the flat rotation velocity, called the baryonic Tully-Fisher relation
 \citep[BTFR,][]{McGaugh00, Verheijen01, McGaugh11, Lelli16, Lelli19}.
Similarly in elliptical galaxies and galaxy clusters,
 such a correlation was found between the baryonic mass and the velocity dispersion, called the baryonic Faber-Jackson relation \citep[BFJR,][]{Sanders10, FM12}.

In galaxies, the BTFR and the BFJR can be implied by the radial acceleration relation (RAR)
 in the low acceleration approximation.
Not long ago, \cite{McGaugh16} have explored a tight correlation between the observed acceleration $\gobs=|\partial\Phi_{\mathrm{obs}}/\partial\,r|=v^{2}/r$ and
 the baryonic acceleration $\gbar=G\Mbar(<r)/r^{2}$ as
\begin{equation}
 \label{eq:RAR}
 \gobs\simeq\sqrt{\gbar\gdag}\,,
\end{equation}
  with an acceleration scale $\gdag=(1.20\pm0.02)\times 10^{-10}$\,m\,s$^{-2}$.
In addition, the RAR implied $v^4=G\Mbar\gdag$ for the BTFR and
 $\sigma^4\simeq G\Mbar\gdag$ for the BFJR \citep{Lelli17, McGaugh20, Milgrom20}.
Particularly, those correlations were clearly described
 in Modified Newtonian Dynamics \citep[MOND,][]{Milgrom83} about four decades ago.
Furthermore, MOND has been tested in gravitational lensing effects \citep{Tian09, TK19, Brouwer21}.
However, MOND found a residual missing mass in galaxy clusters \citep{Sanders03, FM12},
 which questioned the validity of the RAR on the cluster scale.

Recently, \cite{Tian20} investigated the existence of a radial acceleration relation on cluster scales
 using the Cluster Lensing and Supernova survey with Hubble \citep[CLASH,][]{Postman12}.
In the CLASH sample, $\gobs$ was measured
 by strong-lensing, weak-lensing shear-and-magnification data \citep{Umetsu16},
 while $\gbar$ was calculated with X-ray data sets \citep{Donahue14}
 plus the estimated stellar mass from the empirical gas fraction \citep{Chiu18}.
The result is a parallel RAR (CLASH RAR):
\begin{equation}
 \label{eq:CLASHRAR}
 \gobs\simeq\sqrt{\gbar\gddag}\,,
\end{equation}
 with a larger acceleration scale $\gddag=(2.0\pm0.1)\times10^{-9}$\,m\,s$^{-2}$
 than observed in rotating galaxies.
The CLASH RAR is a tight correlation
 with a small intrinsic scatter $\sigma_{\mathrm{int}}=15\percent$.
Other works also demonstrate a similar acceleration scale in galaxy clusters \citep{Pradyumna21}.

While supposing the validity of the CLASH RAR in dynamics,
 one can derive mass-velocity dispersion relation (MVDR), $\sigma^{4}\simeq G\Mbar\gddag$,
 on the BCG-cluster scale.
Moreover, its implications include
 a flat velocity dispersion profile in both BCGs and clusters.
This similarity is reminiscent of the RAR in galaxies,
 albeit with a larger acceleration scale $\gddag$.

Recently, \cite{Tian21} have quantified the MVDR with 29 galaxy clusters
 in the HIghest X-ray FLUx Galaxy Cluster Sample (HIFLUGCS).
In their studies, the baryonic mass was dominated by X-ray gas \citep{Zhang11}
 plus the stellar mass estimated on the scaling relation \citep{Giodini09}.
All 29 HIFLUGCS clusters illustrated a flat tail in the line-of-sight (los) velocity dispersion profile.
By Bayesian statistics, \cite{Tian21} have obtained the MVDR in galaxy cluster as
\begin{equation}
 \label{eq:MVDR_cluster}
 \log\left(\frac{\Mbar}{\Msun}\right)=4.1^{+0.4}_{-0.4}\log\left(\frac{\sigma_{\mathrm los}}{\mathrm{km}\,\mathrm{s}^{-1}}\right)+1.6^{+1.0}_{-1.3}\,,
\end{equation}
 with a small intrinsic scatter of $\sigma_{\mathrm{int}}={12^{+3}_{-3}}\percent$.
In addition, this intercept implied a consistent acceleration scale $\gddag$ of the CLASH RAR.
Consequently, the success of an MVDR in galaxy clusters
 raises the same issue in BCGs.

Although the kinematic scaling relations were widely studied in BCGs
 such as the Faber-Jackson relation and the fundamental plane
 \citep[e.g., see][]{Oegerle91, Zaritsky06, Bernardi07, Samir20},
 it has never been clear if there is continuity in a single MVDR from the scale
 of individual BCGs to clusters of galaxies \citep{Sanders94}.
As for the flat profiles of BCGs,
 it has been a lack of systematic studies of the velocity dispersion profiles.
Besides, there is only one BCG per cluster, and sufficiently sensitive,
 spatially resolved observations have been rare until now.

In this work, we investigate the MVDR with 54 MaNGA BCGs and compare them with 29 HIFLUGCS clusters.
The paper is organized as follows.
In section \ref{sec:Data}, we introduce the data in the MaNGA BCG sample and
 the method to analyze the velocity dispersion profile.
In section \ref{sec:Results}, we inspect the MVDR on the BCG-cluster scale by Bayesian statistics
 and calculate the residuals against four galaxy-cluster properties.
In section \ref{sec:Discussions}, we discuss our results with the consistency of the CLASH RAR
 and the implications for the dark matter problems.
Finally, we summarize our work in section \ref{sec:Summary}.
Throughout this paper, we assume a flat $\Lambda$ cold dark matter ($\Lambda$CDM)
 cosmology with $\Omega_\mathrm{m}=0.3$, $\Omega_\Lambda=0.7$,
 and a Hubble constant of $H_0 = 70$\,km\,s$^{-1}$\,Mpc$^{-1}$.

%
%

\section{Data and Methods}\label{sec:Data}
To study the MVDR in BCGs, we estimated the baryonic mass and
 analyzed their velocity dispersion profiles at the outermost radius.
The baryonic mass is composed of the stellar mass and the gas mass of BCGs.
Since BCGs are elliptical galaxies, stars are usually concentrated in their center.
So, it is difficult to measure their velocity far from the center.

Investigating the internal kinematics structure has motivated
 some observations with IFUs, which employed spatially resolved spectroscopy.
Among them, MaNGA is the largest survey targeted with an unprecedented sample
 of 10,000 nearby galaxies \citep{Bundy15}.

In this work, we exploited BCGs in public MaNGA Product Lunch-6 (MPL-6),
 which contained a total number of 4824 currently.
This data set corresponds to SDSS Data Release Fifteen
 \citep[DR15,][]{Aguado19}.
By combining both data of DR15 and MPL-6,
 we got sufficient information of the baryonic mass and internal kinematics profile
 for this work.

 \subsection{BCG samples}
Because BCGs are the brightest galaxies usually positioned at the center of a galaxy cluster,
 they can be identified by given memberships.
The BCG samples of SDSS has been systematically explored in \cite{Yang07}.
They obtained lots of groups
 by developing the halo-based group finder in Data Release Seventh (DR7).
When matching Yang's catalog with MPL-6,
  73 samples in MaNGA were re-identified as BCGs
  by the color-magnitude and the membership distribution diagrams.
In total, we got 76 BCGs when including three additional samples labeled by MaNGA.

In our studies, both complete spectra and IFU measurements are necessary for calculating
 the baryonic mass and the flat velocity dispersion.
Among the samples, six out of them lack sufficient MaNGA IFU data
 and 16 cases are without the SDSS spectra.
Therefore, only 54 MaNGA BCGs meet the specific requirements of the MVDR.
Their galactic properties are listed in Table~\ref{tab:BCGs}, such as ID, redshift, baryonic mass $\Mbar$,
 and the flat los velocity dispersion $\sigma_{\mathrm{los}}$.

\begin{deluxetable}{lccl}[!htb]
        \tablecolumns{4}
        \tabletypesize{\footnotesize}
        \tablewidth{0.9\columnwidth}
        \tablecaption{\label{tab:BCGs}
        Properties of 54 MaNGA BCGs
        }
        \tablehead{
         \multicolumn{1}{c}{plateifu} &
         \multicolumn{1}{c}{$z$\tablenotemark{a}} &
         \multicolumn{1}{c}{$\log(M_\mathrm{bar})$\tablenotemark{b}} &
         \multicolumn{1}{c}{$\sigma_\mathrm{los}$\tablenotemark{c}} \\
         \colhead{}
         &  &
         \multicolumn{1}{c}{($M_{\odot}$)} &
         \multicolumn{1}{c}{(km/s)}
         }
         \startdata		
8625-12704$^{*}$	&	0.027	&	11.541	$\pm	\,	0.077	$	&	298	$\pm	\,	5	$	\\
9181-12702$^{*}$	&	0.041	&	11.622	$\pm	\,	0.086	$	&	266	$\pm	\,	10	$	\\
9492-9101	&	0.053	&	11.785	$\pm	\,	0.085	$	&	291	$\pm	\,	10	$	\\
8258-3703	&	0.059	&	11.246	$\pm	\,	0.096	$	&	190	$\pm	\,	17	$	\\
8331-12701	&	0.061	&	11.231	$\pm	\,	0.086	$	&	195	$\pm	\,	11	$	\\
8600-12703$^{*}$	&	0.061	&	11.332	$\pm	\,	0.086	$	&	220	$\pm	\,	11	$	\\
8977-3703$^{*}$	&	0.074	&	11.475	$\pm	\,	0.086	$	&	246	$\pm	\,	9	$	\\
8591-3704	&	0.075	&	11.506	$\pm	\,	0.106	$	&	281	$\pm	\,	9	$	\\
8591-6102	&	0.076	&	11.435	$\pm	\,	0.089	$	&	262	$\pm	\,	13	$	\\
8335-6103	&	0.082	&	11.575	$\pm	\,	0.086	$	&	313	$\pm	\,	13	$	\\
9888-12703	&	0.083	&	11.542	$\pm	\,	0.089	$	&	300	$\pm	\,	12	$	\\
9043-3704$^{*}$	&	0.084	&	11.473	$\pm	\,	0.091	$	&	226	$\pm	\,	10	$	\\
9042-3702	&	0.084	&	11.050	$\pm	\,	0.604	$	&	251	$\pm	\,	18	$	\\
8943-9102$^{*}$	&	0.085	&	11.442	$\pm	\,	0.087	$	&	135	$\pm	\,	33	$	\\
8613-6102	&	0.086	&	11.807	$\pm	\,	0.083	$	&	331	$\pm	\,	7	$	\\
9002-3703	&	0.088	&	11.171	$\pm	\,	0.092	$	&	203	$\pm	\,	11	$	\\
8939-6104	&	0.088	&	11.309	$\pm	\,	0.086	$	&	271	$\pm	\,	32	$	\\
8455-12703$^{*}$	&	0.092	&	11.559	$\pm	\,	0.091	$	&	206	$\pm	\,	9	$	\\
9025-9101	&	0.096	&	11.813	$\pm	\,	0.077	$	&	277	$\pm	\,	24	$	\\
8239-6103	&	0.097	&	11.361	$\pm	\,	0.088	$	&	236	$\pm	\,	32	$	\\
9486-6103	&	0.098	&	11.401	$\pm	\,	0.078	$	&	241	$\pm	\,	5	$	\\
8613-12705	&	0.099	&	12.029	$\pm	\,	0.120	$	&	377	$\pm	\,	129	$	\\
9000-9101	&	0.105	&	11.225	$\pm	\,	0.088	$	&	202	$\pm	\,	10	$	\\
8447-3702	&	0.109	&	11.596	$\pm	\,	0.106	$	&	250	$\pm	\,	19	$	\\
9891-9101	&	0.111	&	11.609	$\pm	\,	0.101	$	&	261	$\pm	\,	16	$	\\
8466-6104	&	0.113	&	11.552	$\pm	\,	0.087	$	&	210	$\pm	\,	14	$	\\
9891-3701	&	0.114	&	11.462	$\pm	\,	0.091	$	&	212	$\pm	\,	5	$	\\
8081-3701	&	0.115	&	11.556	$\pm	\,	0.094	$	&	281	$\pm	\,	9	$	\\
9044-12703	&	0.117	&	11.414	$\pm	\,	0.620	$	&	316	$\pm	\,	10	$	\\
8131-3703	&	0.119	&	11.592	$\pm	\,	0.084	$	&	272	$\pm	\,	27	$	\\
9891-12701	&	0.120	&	11.442	$\pm	\,	0.095	$	&	323	$\pm	\,	67	$	\\
9085-6102	&	0.120	&	11.702	$\pm	\,	0.092	$	&	307	$\pm	\,	13	$	\\
9506-6103$^{*}$	&	0.120	&	11.578	$\pm	\,	0.087	$	&	297	$\pm	\,	13	$	\\
8943-3704	&	0.124	&	11.697	$\pm	\,	0.082	$	&	266	$\pm	\,	34	$	\\
9490-9102	&	0.125	&	11.619	$\pm	\,	0.088	$	&	278	$\pm	\,	14	$	\\
9043-9101	&	0.127	&	11.582	$\pm	\,	0.092	$	&	336	$\pm	\,	35	$	\\
8725-12704	&	0.129	&	12.068	$\pm	\,	0.086	$	&	504	$\pm	\,	56	$	\\
8989-12704	&	0.129	&	11.758	$\pm	\,	0.094	$	&	285	$\pm	\,	39	$	\\
8989-12703	&	0.130	&	11.644	$\pm	\,	0.124	$	&	279	$\pm	\,	24	$	\\
8728-3703	&	0.131	&	11.596	$\pm	\,	0.090	$	&	226	$\pm	\,	23	$	\\
9865-12703	&	0.131	&	11.498	$\pm	\,	0.083	$	&	211	$\pm	\,	8	$	\\
8717-1901	&	0.131	&	11.493	$\pm	\,	0.081	$	&	268	$\pm	\,	3	$	\\
8991-6102	&	0.133	&	11.654	$\pm	\,	0.099	$	&	270	$\pm	\,	7	$	\\
8555-3702	&	0.133	&	11.457	$\pm	\,	0.089	$	&	240	$\pm	\,	19	$	\\
8554-6103	&	0.133	&	11.594	$\pm	\,	0.088	$	&	297	$\pm	\,	57	$	\\
8995-6103	&	0.133	&	11.458	$\pm	\,	0.096	$	&	208	$\pm	\,	19	$	\\
8720-12705	&	0.135	&	11.236	$\pm	\,	0.096	$	&	228	$\pm	\,	51	$	\\
8615-12704	&	0.135	&	11.572	$\pm	\,	0.081	$	&	471	$\pm	\,	95	$	\\
8554-6102	&	0.136	&	11.775	$\pm	\,	0.115	$	&	304	$\pm	\,	12	$	\\
8616-12703	&	0.138	&	11.884	$\pm	\,	0.089	$	&	308	$\pm	\,	11	$	\\
8616-3702	&	0.138	&	11.581	$\pm	\,	0.089	$	&	284	$\pm	\,	13	$	\\
8247-9102$^{*}$	&	0.140	&	11.734	$\pm	\,	0.084	$	&	330	$\pm	\,	24	$	\\
9888-6104	&	0.147	&	11.729	$\pm	\,	0.087	$	&	265	$\pm	\,	6	$	\\
8725-6104	&	0.148	&	11.614	$\pm	\,	0.083	$	&	270	$\pm	\,	38	$	
\enddata
\tablenotetext{}{Notes.}
\tablenotetext{a}{Redshifts are from MaNGA Pipe3D.}
\tablenotetext{b}{The baryonic mass including total stellar mass estimated by model photometry in SDSS DR15 and the measured gas mass in MaNGA marked with $*$ on the plateifu ID.}
\tablenotetext{c}{The flat los velocity dispersion from MaNGA IFU in this work.}
\end{deluxetable}

  \begin{figure*}[!htb]
  \includegraphics[width=2.0\columnwidth]{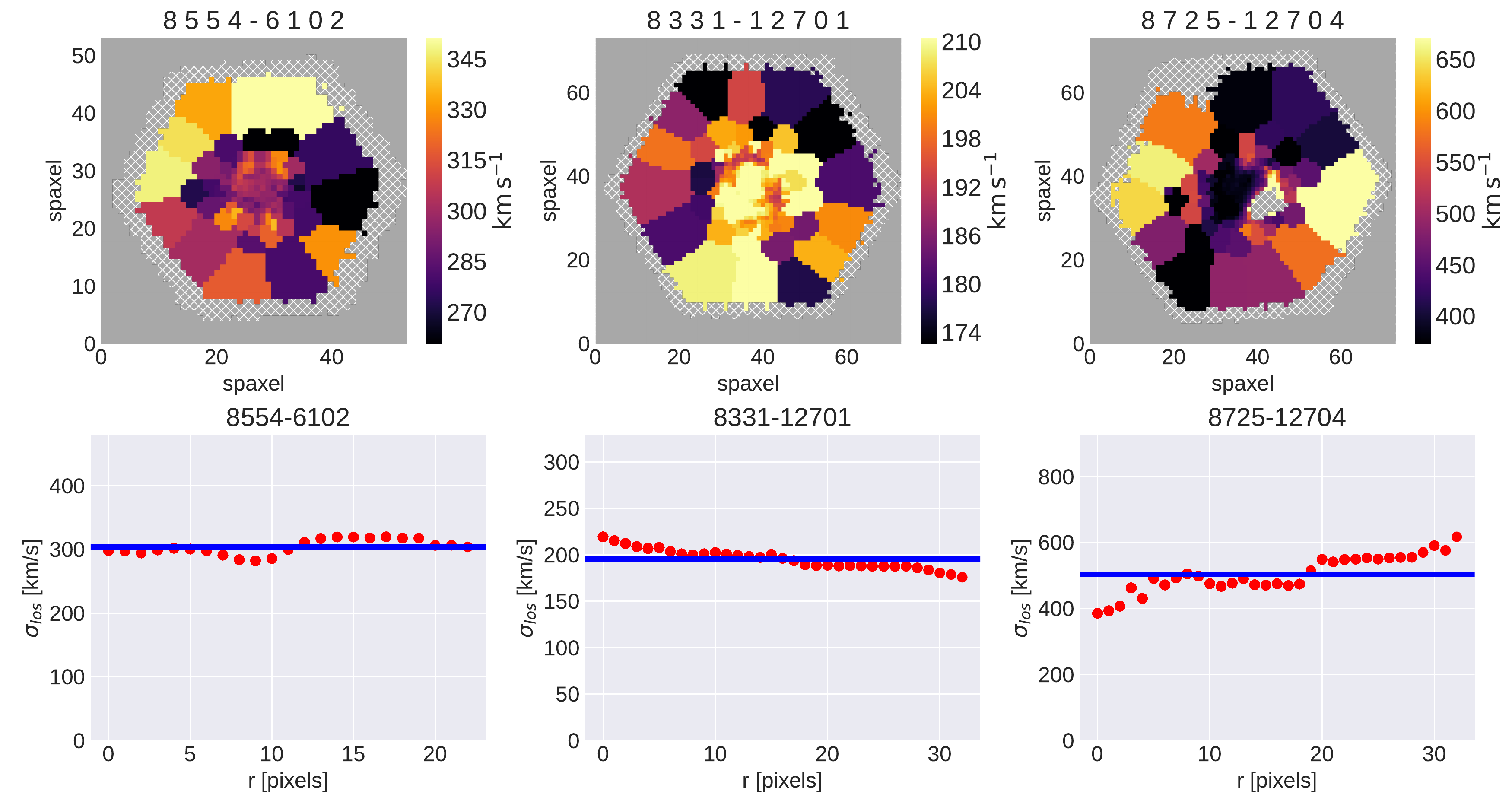}
  \caption{
Three examples of MaNGA BCGs with the plateifu of
 `8554-6102', `8331-12701', and `8725-12704'.
Upper panel: The map plot of \texttt{Spaxel} data for the stellar velocity dispersion.
Lower panel: velocity dispersion profiles in terms of radius.
The red circles represent the los velocity dispersion of concentric circles
 at different radius.
The blue solid lines represents the mean of red data points,
  which is adopted as the flat los velocity dispersion of each BCG.
  }\label{fig:1}
  \end{figure*}

    \subsection{The Baryonic Mass}
Because BCGs are usually elliptical galaxies,
 the baryonic mass is dominated by their stellar mass.
The total stellar mass of 54 MaNGA BCGs can be estimated by model photometry in SDSS DR15.
We adopt uncertainties that average over the occasional asymmetry
 in the photometric stellar mass estimates.
The gas mass of BCGs is subdominant to stars.
Only nine galaxies have measured gas masses in MaNGA
 (e.g., see the star marker $*$ of plateifu ID in Table~\ref{tab:BCGs}),
 with the gas fraction in our samples ranging
 from $0.03\percent-3.2\percent$ of the baryonic mass.
Moreover, the hot gas contribution in the inner region
 (within the effective radius $R_{\mathrm{e}}$)
 is insignificant compared to the stellar mass of BCG \citep[e.g.,][]{Sartoris20, Tian20}.

    \subsection{The Flat Velocity Dispersion}
The flat los velocity dispersion of BCGs can be calculated
 by its one-dimensional profile relative to their centers.
We apply \marvin~developed in \textit{Python} \citep{Cherinka19}
 to access and analyze MaNGA \spaxel~data for the los stellar velocity dispersion
 (Figure~\ref{fig:1}).
All data below 20 km s$^{-1}$ are discarded, because those are considered
 unreliable \citep{Bundy15, Durazo18}.
We calculate the mean los velocity dispersion of each circle.
We find the velocity dispersion profiles of BCGs to be remarkably flat.
Only one BCG in our sample `8943-9102' exhibits a strongly declining profile.
We therefore adopt the average $\sigma_{\mathrm los}$ as the characteristic value
 for each galaxy (blue lines in Figure~\ref{fig:1}).

%
%

\section{Results}\label{sec:Results}
Our main goal is to explore the empirical kinematic scaling relation
 between two independent measurements in BCGs:
 the baryonic mass and the flat los velocity dispersion of stellar components.
In addition to the tightness and the correlations among the galactic properties,
 we also study the intrinsic scatter and the residuals with Bayesian statistics.
The relation and the residuals are elucidated in the following subsections.

\subsection{MVDR with Bayesian Statistics}
  \begin{figure*}[!htb]
  \includegraphics[width=0.80\columnwidth]{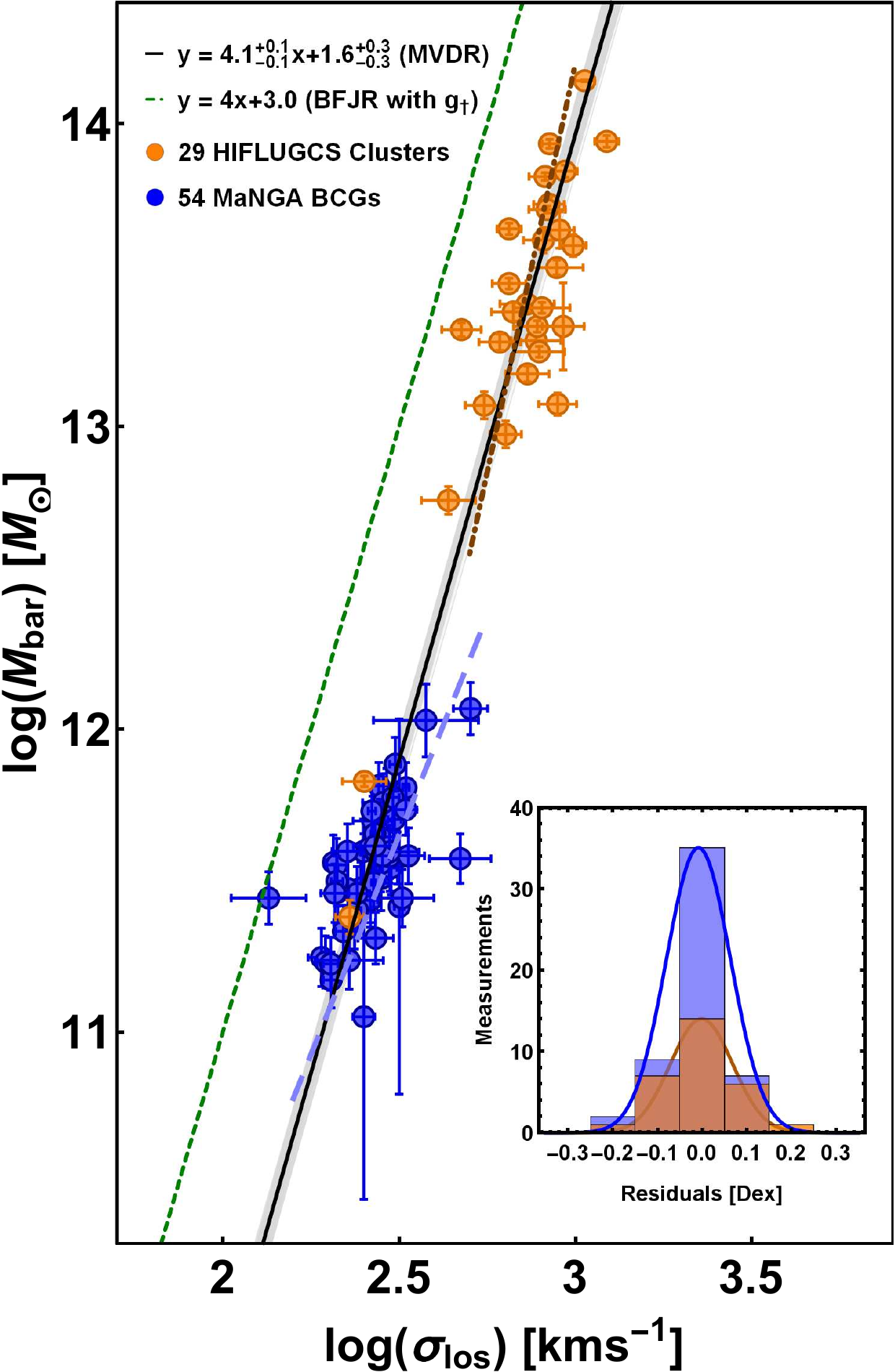}
  \includegraphics[width=1.15\columnwidth]{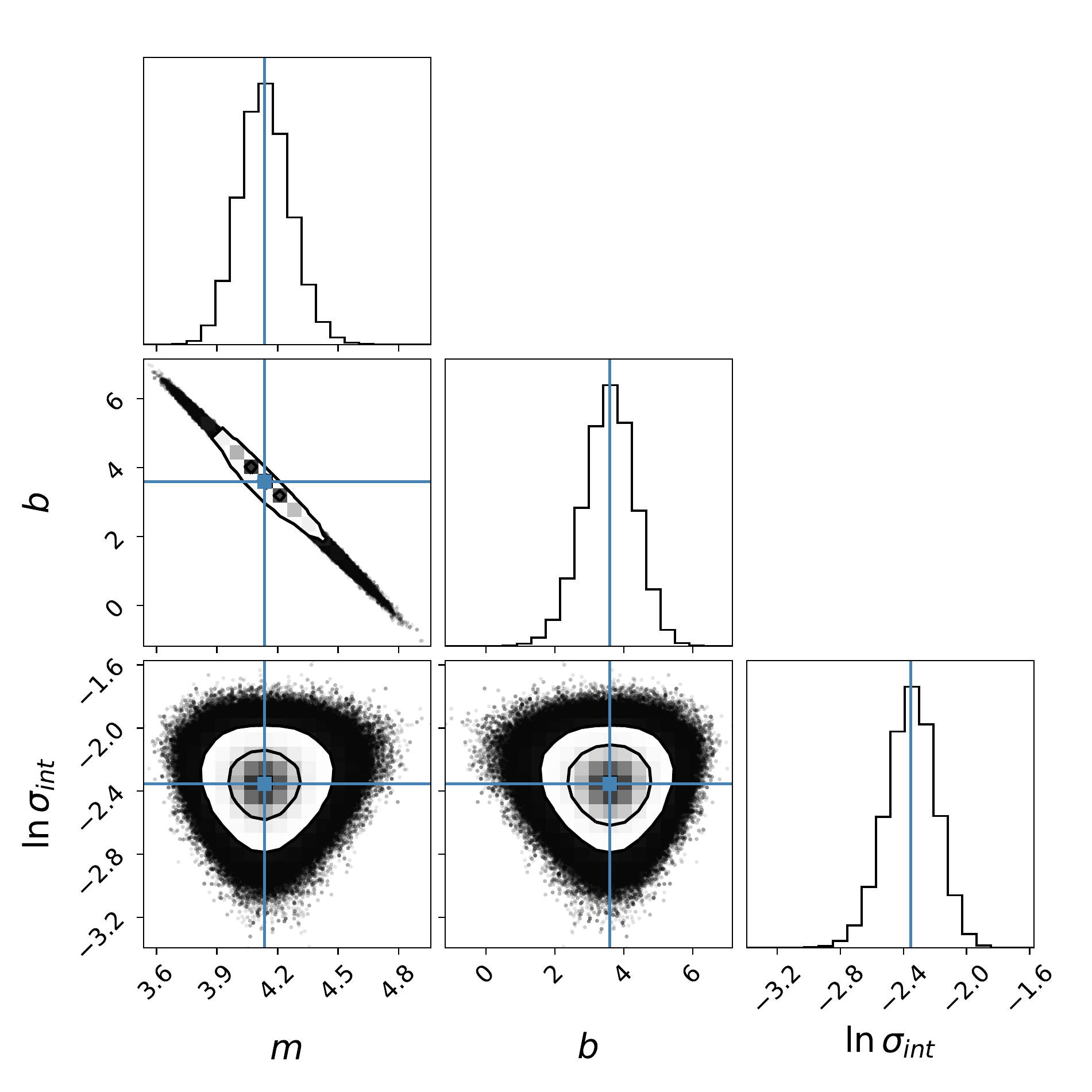}
  \caption{The MVDR of both BCGs and clusters.
Left panel:
The blue circles represent 54 MaNGA BCGs while
 the orange circles indicate 29 HIFLUGCS clusters in \cite{Tian21}.
The black solid line represent the MVDR of all samples,
 $\log (\Mbar/\Msun)=4.1^{+0.1}_{-0.1}\log(\sigma_{\mathrm los}/\mathrm{km}\,\mathrm{s}^{-1})+1.6^{+0.3}_{-0.3}$\,.
The gray shaded area illustrates the one-sigma error of the best fit.
The inset panel demonstrates the histograms of the orthogonal residuals of
 both BCGs (blue) and clusters (orange).
The histograms of both samples indicate Gaussian distributions with
 the same tiny half-width of 0.07 dex.
The fitting results of a narrow dynamical range are illustrated with
 the brown dot-dashed line for 27 HIFLUGCS clusters without two smallest samples and
 the blue long dashed line for 54 MaNGA BCGs, respectively.
The slopes of these lines are uncertain for lack of dynamic range;
 the BCG and cluster samples clearly connect when considered together.
For comparison, the green dashed line demonstrates the BFJR with the acceleration scale of $\gdag$
 assuming Jeans factor $J_{\infty}=16$, $\log (\Mbar/\Msun)=4\log(\sigma_{\mathrm los}/\mathrm{km}\,\mathrm{s}^{-1})+3.0$.
Right panel: Triangle diagrams of the regression parameters for Equation~(\ref{eq:MVDR})
 with marginalized one-dimensional (histograms) and two-dimensional posterior distributions.
Black contours represent 1$\sigma$ and 2$\sigma$ confidence region.
  }\label{fig:2}
  \end{figure*}

In the logarithmic plane of the MVDR,
 54 MaNGA BCGs are distributed as a linear relation.
We model them by introducing $y=mx+b$ with two independent variables:
$y\equiv\ln(\Mbar/\Msun)$
 and $x\equiv\ln(\sigma_{\rm los}/\mathrm{km}\,\mathrm{s}^{-1})$.
With Bayesian statistics,
 we implement a Markov Chain Monte Carlo (MCMC) analysis with
 the orthogonal-distance-regression (ODR) method suggested in \cite{Lelli19, Tian21}.

    The log-likelihood function is written as
\begin{equation}\label{eq:log-likelihood}
     -2\ln\,\mathcal{L}=\sum_{i}\,\ln{(2\pi\sigma^2_{i})}+\sum_{i}\,\frac{\Delta_i^2}{\sigma^2_{i}}\,,
\end{equation}
    with
\begin{equation}\label{eq:Delta}
    \Delta_i^2=\frac{(y_i-m\,x_i-b)^2}{m^2+1}\,,
\end{equation}
    where $i$ runs over all data points, and $\sigma_i$ includes the
    observational uncertainties $(\sigma_{x_i}, \sigma_{y_i})$ and
    the lognormal intrinsic scatter $\sigma_\mathrm{int}$,
\begin{equation}\label{eq:sigma}
     \sigma^2_{i}=\frac{m^2\sigma^2_{x_i}}{m^2+1}+\frac{\sigma^2_{y_i}}{m^2+1}+\sigma_\mathrm{int}^2\,.
\end{equation}

We perform the ODR MCMC analysis for the slope and the intercept
 implemented in \textit{Python} \citep[\textit{emcee};][]{Foreman-Mackey13, emcee19}.
While employing non-informative flat priors on the slope $m$ and the intercept $b$
 within the interval of $[-100, 100]$, and the intrinsic scatter $\ln(\sigma_{\rm int})\,\in[-5,2]$,
 we discover a tighter correlation by
\begin{equation}\label{eq:MVDR}
    \log\left(\frac{\Mbar}{\Msun}\right)=4.1^{+0.1}_{-0.1}\log\left(\frac{\sigma_{\mathrm los}}{\mathrm{km}\,\mathrm{s}^{-1}}\right)+1.6^{+0.3}_{-0.3}\,,
\end{equation}
 with a tiny error of the lognormal intrinsic scatter of
 $10^{+2}_{-1}\percent$.
We present the regression parameters with the posteriors distribution
 in the right panel of Figure~\ref{fig:2}.

To justify the initial assumption of a Gaussian intrinsic scatter
 perpendicular to the fitting line,
 we examine the histogram of the orthogonal residuals $\Delta_i$ with respect to Equation~(\ref{eq:MVDR})
 (see the inset panel of Figure~\ref{fig:2}).
The distributions of the residuals demonstrate a Gaussian distribution
 with a tiny half-width (0.07 dex) for both BCGs and clusters.

To derive the precise value of the acceleration scale by the intercept,
 we apply a fixed slope ($m=4$) and perform the ODR MCMC method again.
It gives
\begin{equation}\label{eq:MVDR m=4}
    \log\left(\frac{\Mbar}{\Msun}\right)=4\log\left(\frac{\sigma_{\rm los}}{\mathrm{km}\,\mathrm{s}^{-1}}\right)+1.90^{+0.02}_{-0.02}\,,
\end{equation}
 with the same lognormal intrinsic scatter
 $\sigma_{\mathrm{int}}=10^{+2}_{-1}\percent$.
Accordingly, the intercept implies an acceleration scale of
 $\gddag=(0.9-2.4)\times10^{-9}$ m s$^{-2}$
 by assuming Jeans factor $J_{\infty}\in [9, 25]$
 \citep[for detail see, e.g., section 4.1 in][]{Tian21}.
This is consistent with that of the CLASH RAR,
 but larger than observed in spirals \citep{McGaugh16}.

We also consider the fitting results of BCGs and clusters separately.
Two distinct distributions can be divided as 54 MaNGA BCGs
 and 27 HIFLUGCS clusters.
By implementing the same method, we get a different MVDR of BCGs with
 $m=2.9^{+0.5}_{-0.3}$ and $b=10.1^{+1.9}_{-2.6}$.
In addition, the other case has been analyzed before:
 $m=5.4^{+2.2}_{-1.1}$ and $b=-2.0^{+3.2}_{-6.3}$ \citep[e.g., see the section 3.1 in][]{Tian21}.
Two results are illustrated with the blue dashed and the brown dot-dashed line
 in Figure~\ref{fig:2}, respectively.
Conversely, the uncertainty of the slopes and intercepts are much larger than the full samples
 due to the narrow dynamical range.
Consequently, the two relations are consistent with being drawn
 from a single underlying relation spanning the entire range from BCGs to clusters.

To estimate the difference in another common fitting method,
 we calculate the intrinsic scatter along the vertical direction
 \citep[e.g., see appendix A in][]{Lelli19}.
While performing the vertical MCMC method, we find a similar relation as
 $\log (\Mbar/\Msun)=3.9^{+0.1}_{-0.1}\log(\sigma_{\mathrm los}/\mathrm{km}\,\mathrm{s}^{-1})+2.3^{+0.3}_{-0.3}$\,,
 albeit with a much larger lognormal intrinsic scatter
 $\sigma_{\mathrm{int}}=40\pm6\percent$.
Regardless of the scatter, the difference in the slope is negligible between the two methods.

\subsection{Residuals}
\begin{figure*}[!htb]
    \centering
    \includegraphics[width=2.0\columnwidth]{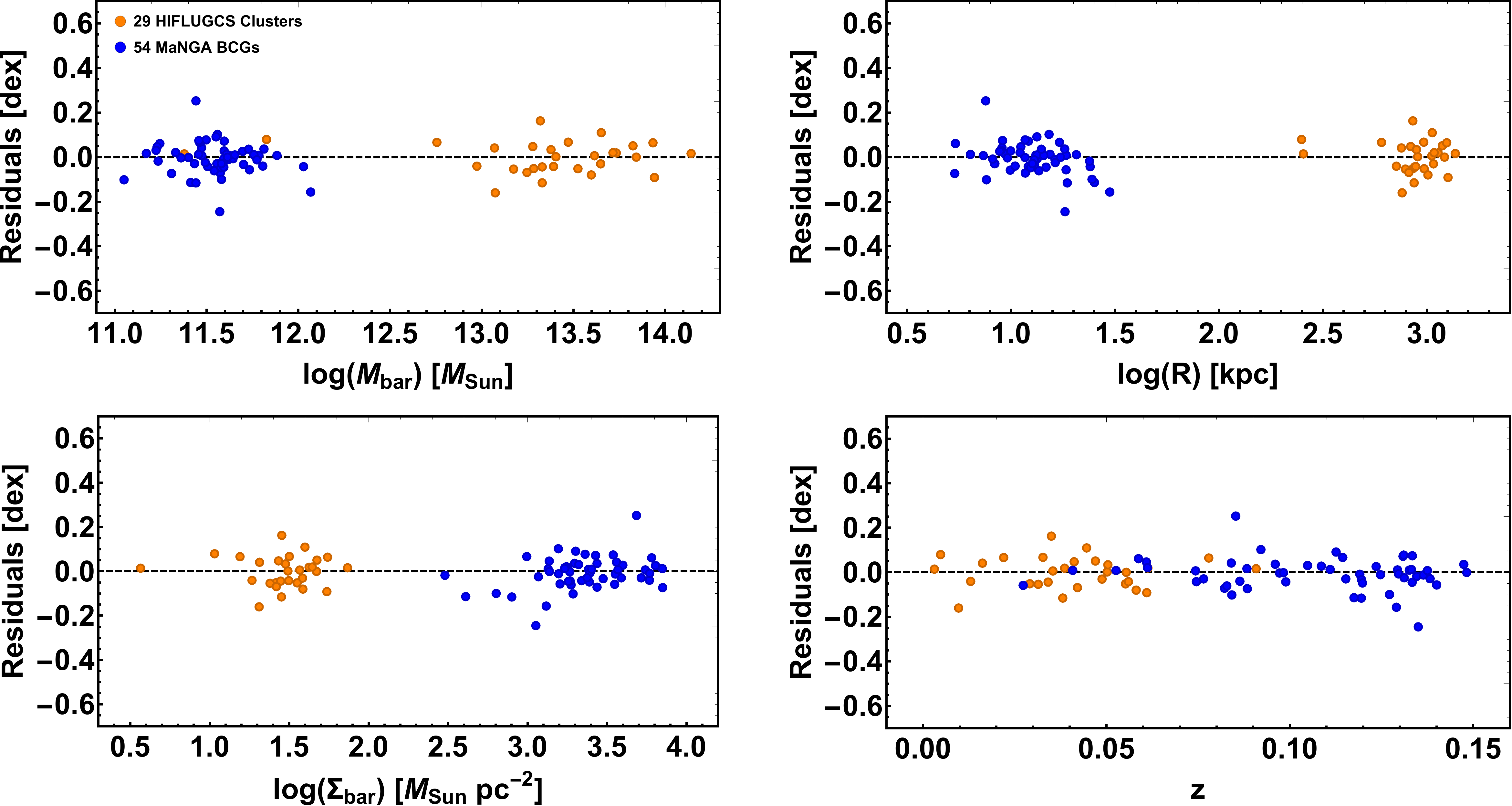}
    \caption{
The orthogonal residuals after subtracting Equation~(\ref{eq:MVDR})
 against four global quantities of galaxies and clusters:
 baryonic mass $\Mbar$ (upper-left), scale length $R_{\mathrm{e}}$ for galaxies; $r_{500}$ for clusters (upper-right), baryonic mass surface density $\Sigma_{\mathrm{bar}}$ (lower-left), and redshift (lower-right).
Blue circles denote BCGs and orange circles denote clusters.
The dashed line represents zero difference.
    }\label{fig:3}
\end{figure*}

To examine the deviation of the dependence on some major properties of galaxies and clusters,
  we consider the orthogonal residuals after subtracting Equation~(\ref{eq:MVDR}) against four global quantities: baryonic mass, scale length ($R_{\mathrm{e}}$ for galaxies; $r_{500}$ for clusters), surface density, and redshift in Figure~\ref{fig:3}.
The residuals of all samples are distributed within a tiny range from $-0.02$ to $0.02$ dex.
Moreover, all diagrams display insignificant correlations with four galaxy-cluster properties,
  which is reminiscent of the BTFR for individual galaxies.
Only the mass and velocity dispersion seem to matter.
There is no obvious second parameter.

%
%

\section{Discussions}\label{sec:Discussions}
For the first time, we reveal a tight empirical kinematic correlation,
 i.e. MVDR, on the BCG-cluster scale.
The MVDR is a counterpart of a dynamical relation rather than a coincidence,
 which can be derived by the CLASH RAR
 \citep[][]{Tian20}.
As a new discovery of a strong correlation,
 the MVDR can provide a crucial test for the dark matter problem.

    \subsection{Consistency with the CLASH RAR}
The CLASH RAR derive three implications of the kinematics in BCGs and clusters \citep{Tian20, Tian21}:
(1) the flat velocity dispersion profile in BCGs;
(2) the flat velocity dispersion profile in galaxy clusters;
(3) the MVDR on the BCG-cluster scale as $\sigma^{4}\propto~G\Mbar\gddag$.
Initially, the implications in galaxy clusters were first confirmed in \cite{Tian21}
 for (2) and partially (3), see Equation~(\ref{eq:MVDR_cluster}).
Subsequently, our studies examine the rest of (1) and (3) in MaNGA BCGs and
 validate all the implications of the CLASH RAR.
In addition, the MVDR on the BCG-cluster scale is identical
 to that in HIFLUGCS clusters, see Equation~(\ref{eq:MVDR}) and Equation~(\ref{eq:MVDR_cluster}).

A larger acceleration scale also determines
 a smaller scale length $r_{\ddagger}$ of a flat velocity dispersion profile.
In pressure supported systems, \cite{Milgrom84} defined a scale parameter as
 $r_{\dagger}=\sigma_{\mathrm{los}}^2\gdag^{-1}$, for a flat velocity dispersion.
When calculating with $\gddag$ instead,
 we found $r_{\ddagger}=(0.4-4.1)$ kpc in MaNGA BCGs.
Moreover, this scale is much smaller than a typical $R_{\mathrm{e}}\approx30$ kpc \citep[e.g., see][]{Tian20},
 which indicates a flat velocity dispersion even in the innermost region of BCGs.

Besides all the success of our results, one exception in our samples needs to be stressed.
BCG `8943-9102' displays a declining profile and contributes the smallest velocity dispersion
 among the sample.
This exception raises an interesting issue on BCGs:
 whether it appertains to a flat profile inherently.

    \subsection{Implications for Dark Matter Problems}
The $\Lambda$CDM model implied the slope of three for the MVDR on the BCG-cluster scale,
 by assuming a constant baryon fraction \citep[e.g., see section 4.2 in][]{Tian21}.
According their studies, the prediction in the $\Lambda$CDM model demonstrated
 a discrepancy for smaller galaxy clusters.
Coincidentally, MaNGA BCGs also disfavors the $\Lambda$CDM model with the same offset.
On the contrary, the BFJR of elliptical galaxies were explained
 by adopting the abundance matching relation \citep[e.g., see the discussions in][]{DW17, Navarro17}.
Regardless, it still remains a mystery of such an explanation on the BCG-cluster scale.

MOND naturally explained the slope of four for the MVDR on the BCG-cluster scale,
 although a larger acceleration scale $\gddag$ needs to be explained.
One conceivable interpretation is the acceleration scale depending
 on the depth of the potential well \citep{ZF12, HZ17}.
Besides, two characteristic scales could imply an underlying phase transition mechanism behind.
Nevertheless, our discoveries provide a strict test for the attempts on
 the fundamental theory in MOND paradigm.

%
%

\section{Summary}\label{sec:Summary}
In the galactic systems, the acceleration scale $\gdag$ dominated in three empirical scaling relations:
 the RAR, the BTFR and the BFJR.
Among them, a tight dynamical relation, the RAR, can infer two kinematic
 counterparts of the BTFR and the BFJR, and vice versa.
In addition, the acceleration scale $\gdag$ determined a scale distance $r_{\mathrm{\dagger}}$
 for a flat rotation curve and a flat velocity dispersion profile.

On the BCG-cluster scale, our studies filled the gap with such empirical scaling relation,
 the MVDR, albeit with a larger acceleration scale $\gddag$.
The MVDR can be implied by the CLASH RAR as its kinematic relation, and vice versa.
While investigating 54 MaNGA BCGs with IFU measurements,
 our works indicated a flat dispersion profile among all samples except one.
Moreover, the flat tail can be determined by $r_{\mathrm{\ddagger}}$
  according to $\gddag$.

In summary, the consistency between the CLASH RAR and the MVDR was confirmed
 by three different data sets: CLASH samples, HIFLUGCS clusters, and MaNGA BCGs.
Both tight empirical correlations demonstrated tiny intrinsic scatters calculated by Bayesian statistics.
Coincidentally, it raises an issue on such similarity between a galaxy and a cluster scale
 with two different acceleration scales: $\gdag$ and $\gddag$, respectively.
Nevertheless, the consistency of these relations provides a strict test for the dark matter problem.

\section*{ACKNOWLEDGMENTS}
We thank Yen-Ting Lin and Shemile Loriega Poblete for the assistance of MaNGA data.
We also thank the anonymous reviewer for valuable comments to improve the clarity of this paper.
YT, HC, and CMK are supported by the Taiwan Ministry of Science and Technology grant
 MOST 109-2112-M-008-005.
YT is also supported in part by grant MOST 110-2112-M-008-015-MY3.
SSM is supported in part by NASA ADAP grant 80NSSC19k0570 and NSF PHY-1911909.
YHH is supported by the Ministry of Science \& Technology of Taiwan under grant MOST 109-2112-M-001-005 and a Career Development Award from Academia Sinica (AS-CDA-106-M01).


\bibliographystyle{yahapj}
\bibliography{reference_BCG}

\end{document}